%% file: prlhww.tex
\documentclass[aps,prl,showpacs,twocolumn,groupedaddress]{revtex4}  

\usepackage{graphicx}
\usepackage{dcolumn}
\usepackage{bm}
\usepackage{amssymb}   

\include{globals_d0}

\begin{document}


\hspace{5.2in} \mbox{FERMILAB-PUB-05/377-E}

\title{Search for the Higgs boson in \hww\ decays
 in \ppbar\ collisions at $\sqrt{s} = 1.96$ TeV}

\input list_of_authors_r2.tex  
\date{August 27, 2005}

\begin{abstract}
We present a search for the standard model Higgs boson 
in \hww\, decays with \ee, \emu\ and \mumu\ final states 
in \ppbar\ collisions at a center-of-mass energy of $\sqrt{s}=1.96$ TeV.
The data, collected from April 2002 to June 2004 with the D\O\ detector,
correspond to an integrated luminosity of 300 - 325 \ipb, 
depending on the final state.
The number of events observed is consistent with the expectation 
from backgrounds. Limits from the combination of all three
channels on the Higgs production cross section 
times branching ratio $\sigma\times BR($\hww$)$ are presented.
\end{abstract}

\maketitle

In the standard model (SM), the hypothetical Higgs boson is crucial to the 
understanding of electroweak symmetry breaking (EWSB) and the mass generation 
of electroweak gauge bosons and fermions. Spontaneous EWSB predicts the 
existence of this neutral scalar particle with mass \mh, a free parameter in 
the SM. Direct searches at the {\sc cern} $e^+e^-$ collider ({\sc lep}) 
yield a lower limit for the Higgs boson mass of 
\mh $>$ 114.4~GeV\,\cite{bib:lephiggs} at the 95\% CL.
Indirect measurements via fits to the electroweak precision data give an
upper bound of $\mh < 280\,{\rm GeV}$\,\cite{bib:lepewg} at the 95\% CL. 

In this Letter, we present a search for the Higgs boson in \hwwll\ 
($\ell,\ell'=e,\mu,\tau$) 
decays with \ee, \emu, or \mumu\ final states. Tau decays are 
detected in their leptonic decay modes to electrons or muons.
This is the first search for the Higgs boson at a hadron collider 
in this decay channel, which plays an important role in the overall
discovery potential of the Higgs boson at the Fermilab Tevatron 
Collider\,\cite{Carena:2000yx}.
We use data collected by the  D\O\ detector between April 2002 and
June 2004 in \ppbar\ collisions at $\sqrt{s} = 1.96$ TeV of the 
Fermilab Tevatron Collider. The integrated luminosities 
are $325\pm 21$ \ipb, $318\pm 21$ \ipb\ and $299\pm 19$ \ipb\ 
for the \ee, \emu\ and \mumu\ channels, respectively. 
The differences in the integrated luminosities for various channels are
primarily due to different trigger conditions.
Next-to-leading order (NLO) calculations\,\cite{bib:higlu, bib:hdecay}
predict the product of the SM Higgs boson production cross section and
the branching ratio $\sigma(p\bar p \to H) \times BR(\hww)$
of 11--250 fb for the Higgs masses between 100 and 200 GeV. The 
dominant contribution to the cross section comes from gluon-gluon fusion.
Extensions of the SM including a fourth fermion family\,\cite{Arik:2001iw} 
predict an enhanced Higgs boson production cross section.

We briefly describe the main components of the D\O\ Run II 
detector \cite{d0det} important to this analysis. The central tracking system 
consists of a silicon microstrip tracker (SMT) and
a central fiber tracker (CFT), both located within a 2.0~T axial magnetic
field. The SMT strips have a typical pitch of 50--80 $\mu$m, and the design 
is optimized for tracking and vertexing over the pseudorapidity range 
$|\eta| < 3$, where $\eta = -\ln{(\tan{\theta \over 2})}$, where $\theta$ is
the polar angle with respect to the proton beam. The system
has a six-barrel longitudinal structure, each with a set of four silicon layers
arranged axially around the beam pipe, interspersed with sixteen radial disks.
The CFT with full coverage for $|\eta| < 1.6$ has eight 
thin coaxial barrels, each supporting two doublets of
overlapping scintillating fibers, one doublet 
parallel to the beam axis, the other alternating by $\pm 3^{\circ}$ relative
to the beam axis.

A liquid-argon/uranium calorimeter surrounds the central tracking system and
consists of a central calorimeter (CC)
covering to $|\eta|$ $\approx 1.1$, and two end cap calorimeters (EC) 
extending coverage to $|\eta| < 4.2$, all housed in separate cryostats
\cite{d0cal}. Scintillators between the CC and EC cryostats provide additional
sampling of showers for 1.1 $< |\eta| <$ 1.4.

The muon system is located outside the calorimeters and consists of a layer
of tracking detectors and scintillation trigger counters inside iron toroid 
magnets which provide a 1.8~T magnetic field, followed by two similar layers 
behind each toroid. Tracking in
the muon system for $|\eta| < 1$ relies on 10~cm wide drift tubes \cite{d0cal},
while 1~cm mini-drift tubes are used for $1 < |\eta| < 2$
\,\cite{Abramov:1998ti}.

The \hww\ candidates are selected by single or di-lepton triggers
using a three-level trigger system.  The first trigger
level uses hardware to select electron candidates based on energy deposition in
the electromagnetic part of the calorimeter and selects muon candidates formed
by hits in two layers of the muon scintillator system. Digital signal
processors at the second trigger level form muon track candidate segments 
defined
by hits in the muon drift chambers and scintillators. At the third level,
software algorithms running on a computing farm and exploiting the full event
information are used to make the final selection of events which are recorded
for offline analysis. 

In the offline analysis, electrons are identified as electromagnetic 
showers in the calorimeter.
These showers are selected by comparing the longitudinal and
transverse shower profiles to those expected of the electrons.
The showers must be isolated, deposit most of their energy in the 
electromagnetic part of  the calorimeter, and pass a likelihood
criterion that includes a spatial track match and,
in the CC region, an $E/p$ requirement, where $E$ is the energy of the
calorimeter cluster and $p$ is the
momentum of the track. All electrons are required to be in the
pseudorapidity range $|\eta| < 3.0$. 
The transverse momentum measurement of the electrons is based on calorimeter 
cell and track information.

Muon tracks are reconstructed from 
hits in the wire chambers and scintillators in the muon system and must 
match a track in the central tracker. 
To select isolated muons, the scalar sum of the transverse momentum of
all tracks other than that of the muon in a cone of ${\cal R} = 0.5$ around the
muon track must be less than 4 GeV, where ${\cal R} =
\sqrt{(\Delta\phi)^2+(\Delta\eta)^2}$ and $\phi$ is the azimuthal angle.
Muon detection is performed over the full coverage of the muon system 
$|\eta| < 2.0$.
Muons from cosmic rays are rejected by requiring a timing 
criterion on the hits in the scintillator layers 
as well as applying restrictions on the position of the
muon track with respect to the primary vertex.  
 
The decay of the two $W$ bosons into electrons or muons results in three
different final states $\ee+X$ ($ee$\ channel), $\emu+X$ ($e\mu$\ channel) and
$\mumu+X$ ($\mu\mu$\ channel), each of which consists of two oppositely
charged isolated leptons with high transverse momentum and large missing 
transverse energy, \etmiss , due to the undetected neutrinos.
The selection criteria for each channel were chosen to minimize the
cross section upper limit on Higgs production expected in the absence of 
signal. To take into account the signal kinematic characteristics that change 
with the Higgs boson mass, \mh, some selection cuts are \mh\ 
dependent\,\cite{Han:1998sp}. 
Six Higgs boson masses from 100~GeV to 200~GeV have been studied.

\begin{figure}[t]
\includegraphics[width=0.49\columnwidth]{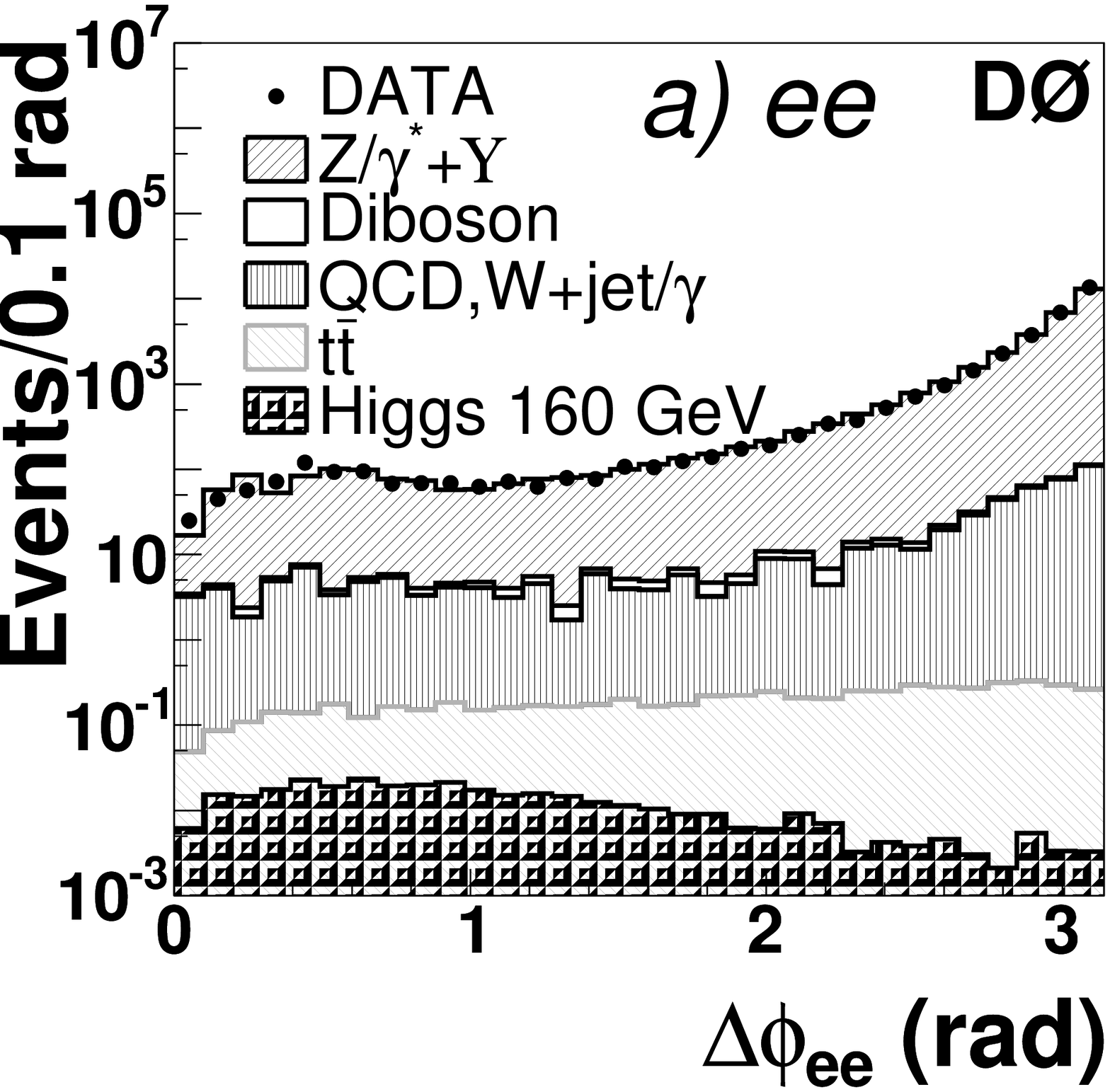}\hfill
\includegraphics[width=0.49\columnwidth]{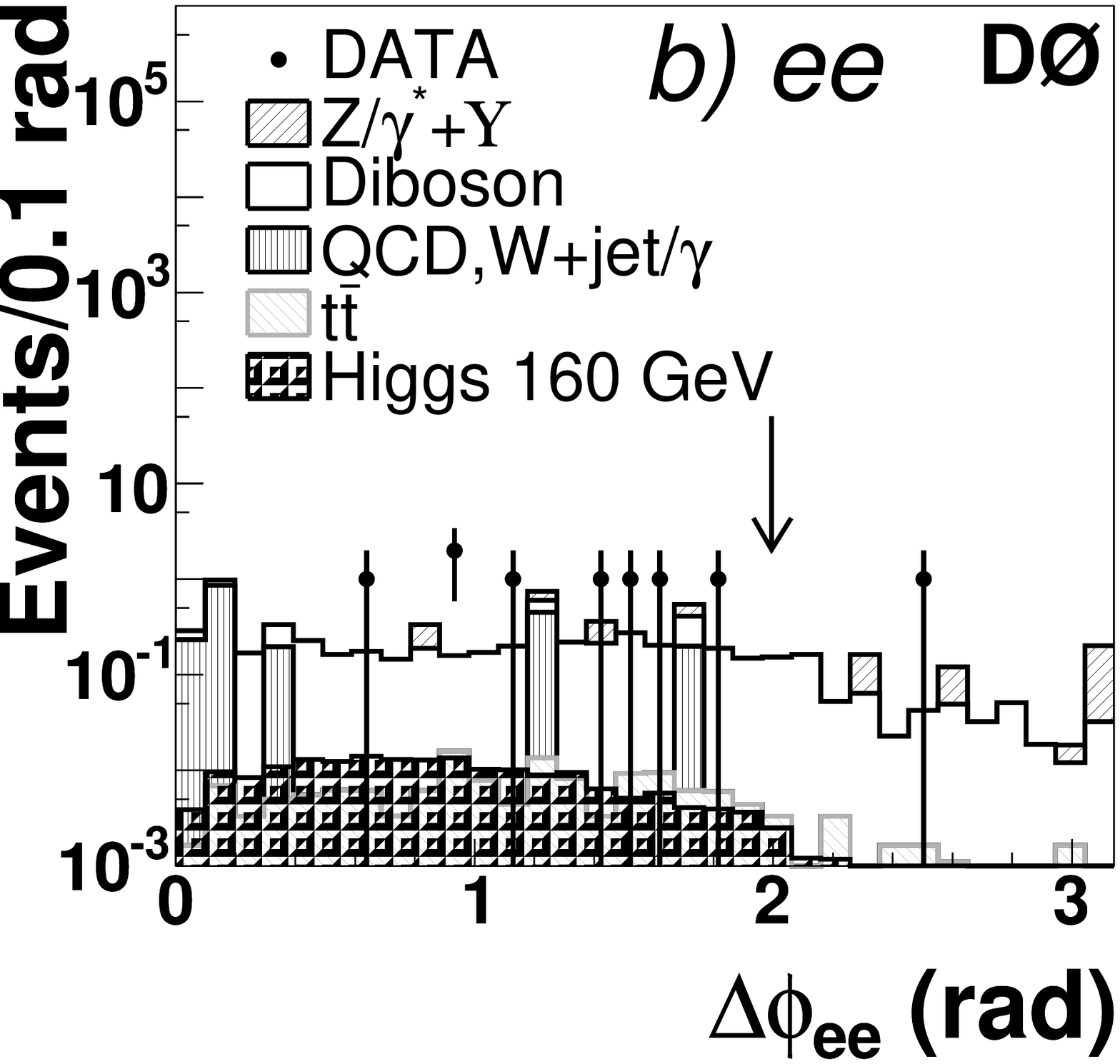}
\includegraphics[width=0.49\columnwidth]{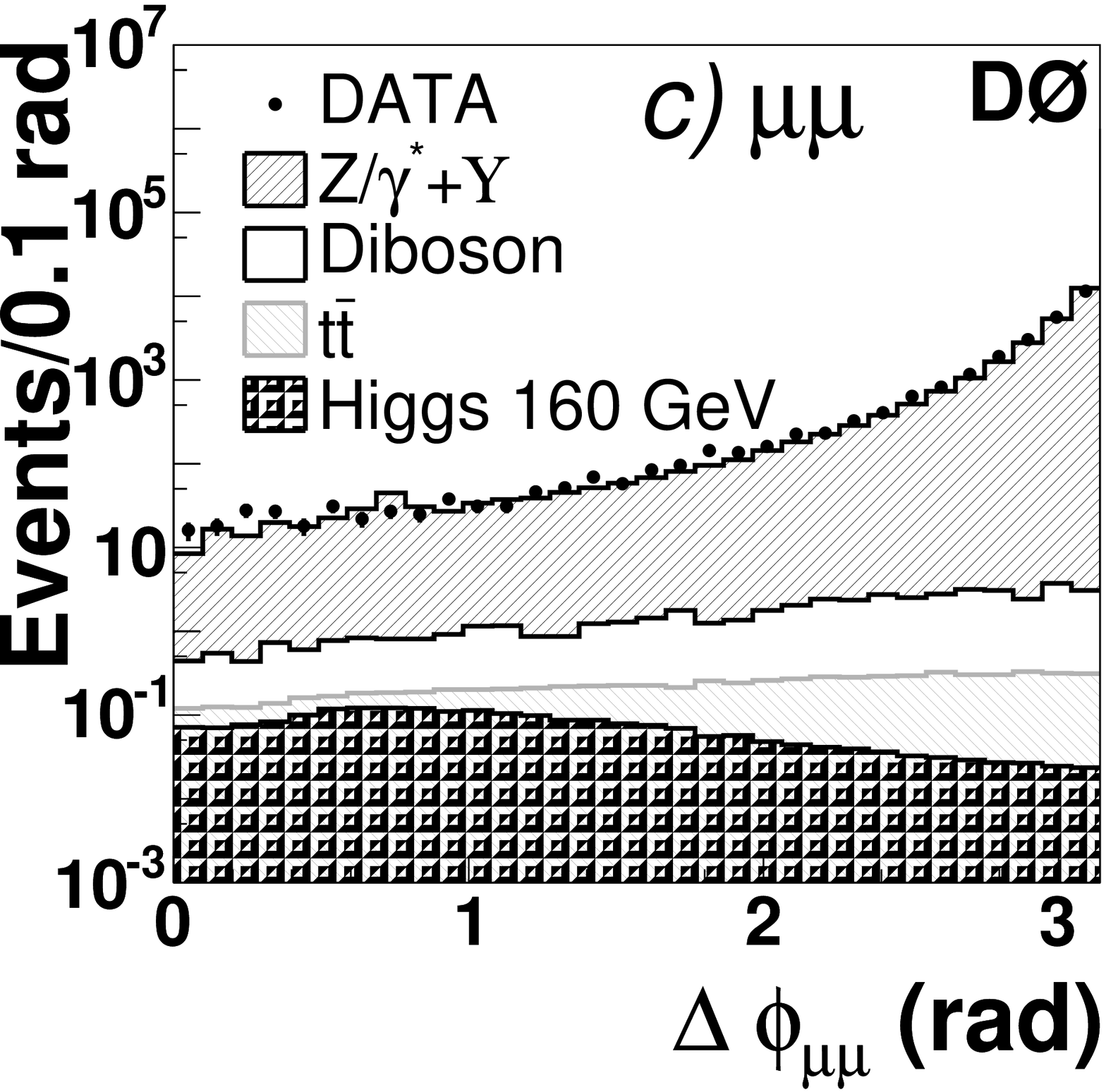}\hfill
\includegraphics[width=0.49\columnwidth]{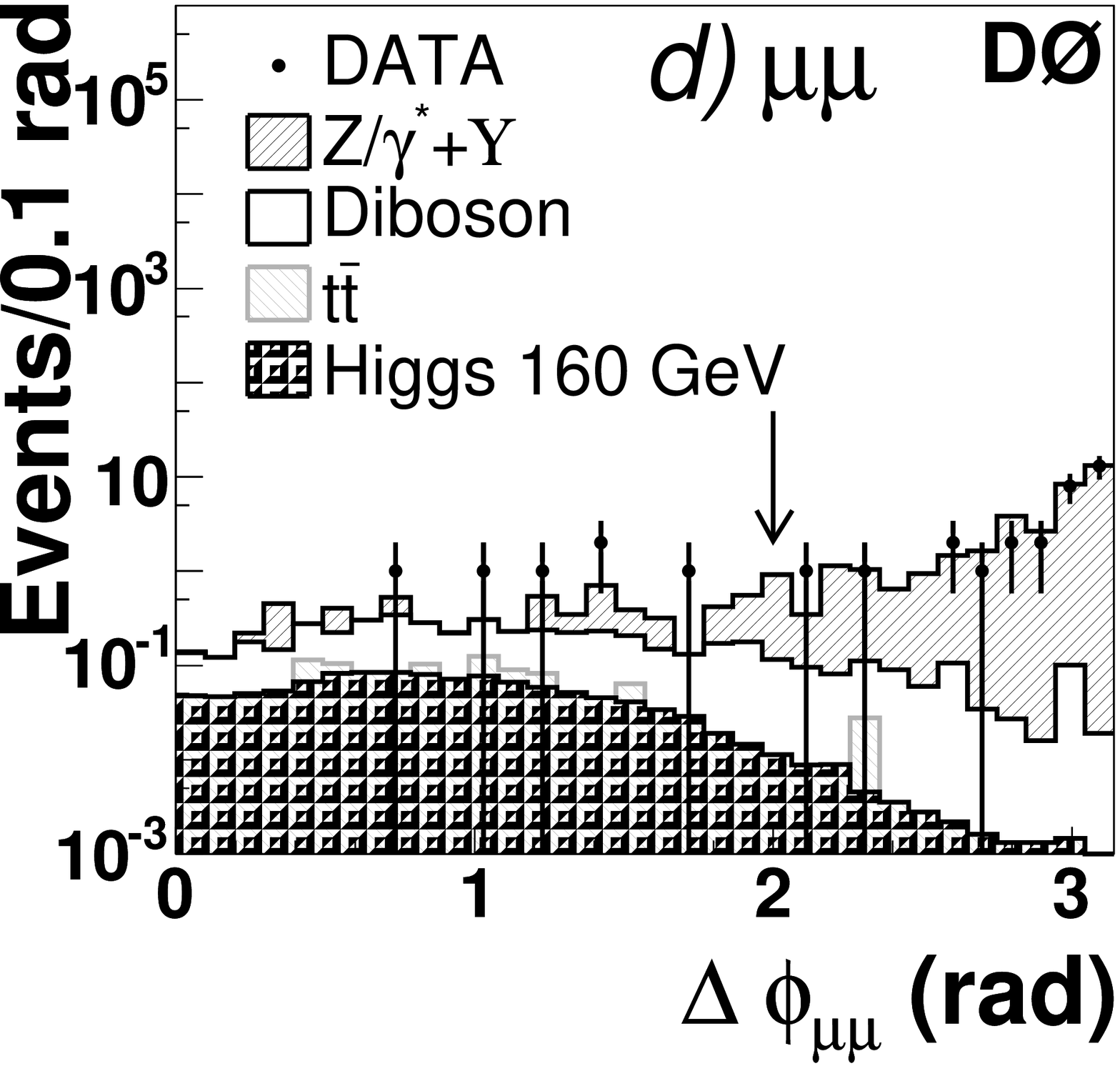}
\includegraphics[width=0.49\columnwidth]{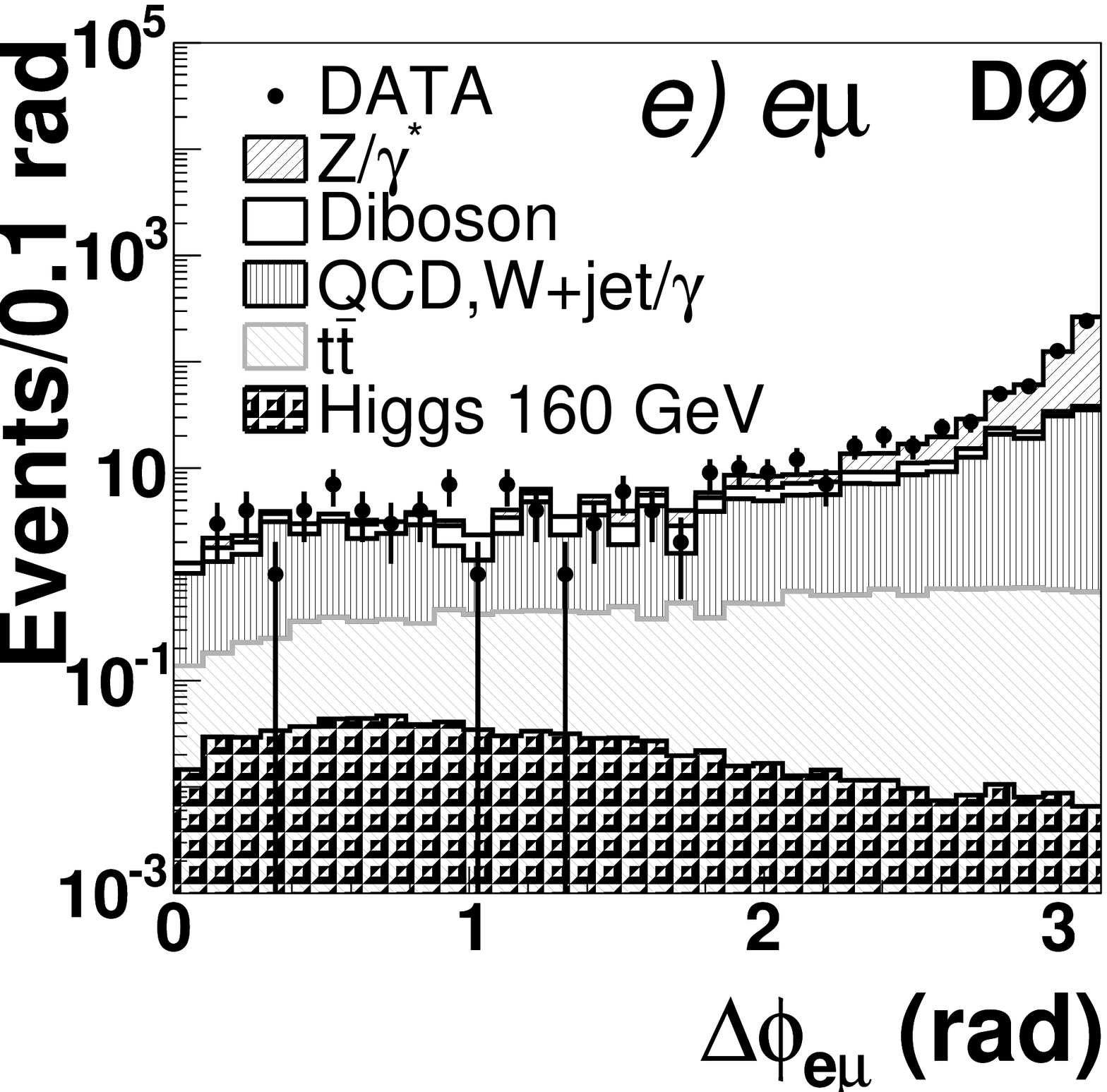}\hfill
\includegraphics[width=0.49\columnwidth]{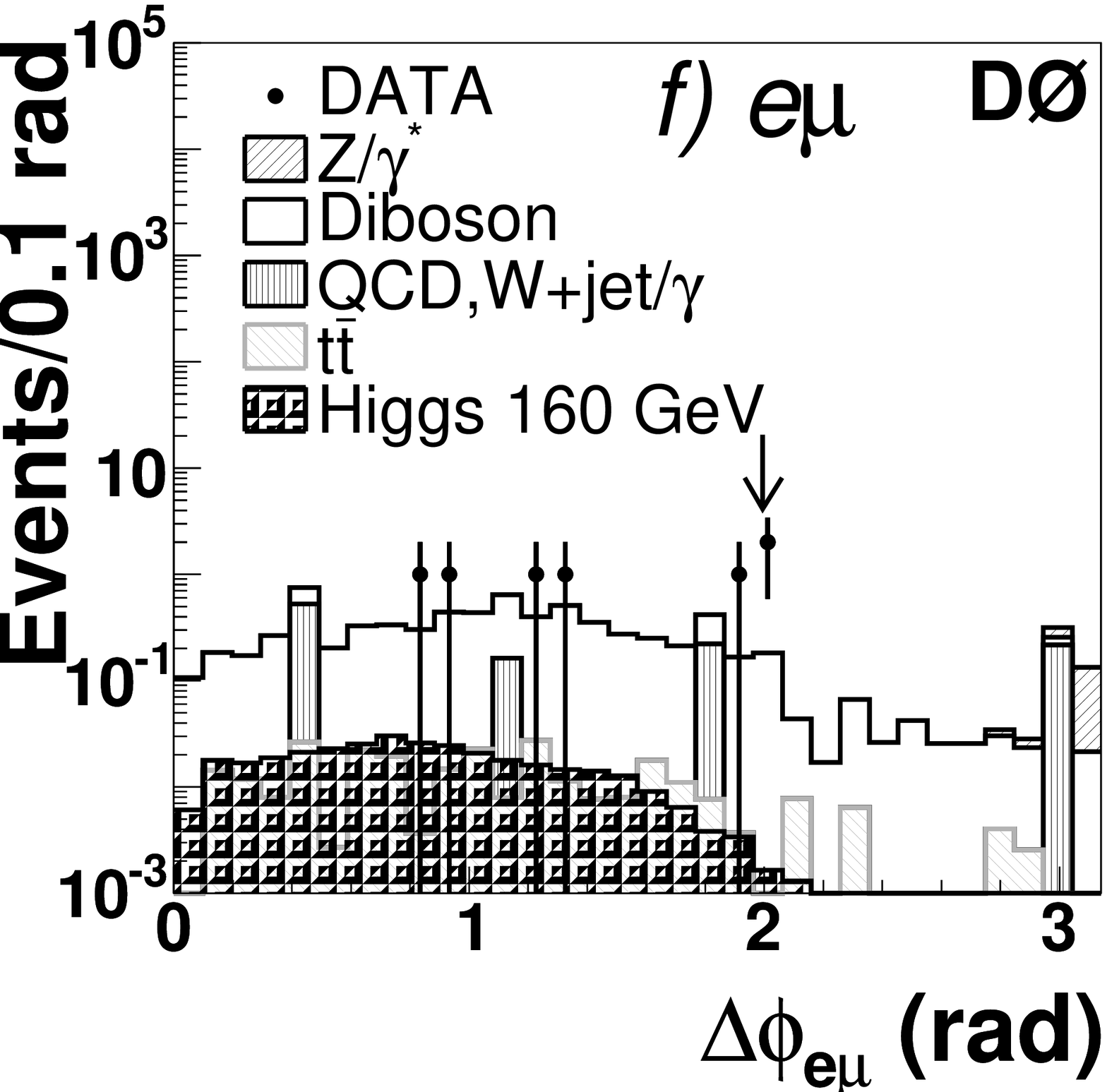}

\caption{\label{fig:dphi_plots} Distribution of the opening angle
  $\Delta \phi_{\ell\ell'}$ after applying the initial transverse 
  momentum cuts in the (a) $ee$, (c) $\mu\mu$ and (e) $e\mu$ channel.
  Figures (b), (d) and (f) show the $\Delta \phi_{\ell\ell'}$ 
  distributions after the final selection except for the 
  $\Delta \phi_{\ell\ell'}$ criterion for the $ee$, $\mu\mu$, and $e\mu$ 
  channel, respectively.
  The arrows indicate the cut values.
  The QCD contribution is negligible in Figs. (c) and (d). }
\end{figure}

In all three channels, two leptons originating from the same vertex
are required to be of opposite charge, and must have transverse momenta
\pt\ $>$ 15~\GeV\ for the leading lepton and \pt\ $>$ 10~\GeV\ for 
the trailing one (Cut 1).
Figure~\ref{fig:dphi_plots} shows the good agreement between data and
Monte Carlo simulation (MC) in distributions of the
azimuthal opening angle $\Delta \phi_{\ell\ell'}$ between the two leptons for 
the $ee$\ (a), the $\mu\mu$\ (c) and the $e\mu$ channel (e) 
after applying the lepton transverse momentum cuts. 

In all cases, the background is largely dominated by $Z/\gamma^*$ production
which is further suppressed by requiring \etmiss\,$>$\,20~GeV in all three 
channels (Cut 2).
Background events are also removed if the \etmiss\ has a large contribution 
from the mis-measurement of jet energy.
The fluctuation in the measurement of jet energy in the transverse plane 
can be approximated by $\Delta E^{\rm jet}\cdot\sin\theta^{\rm jet}$ where
$\Delta E^{\rm jet}$ is proportional to $\sqrt{{E}^{\rm jet}}$. 
The opening angle $\Delta\phi\left({\rm jet},\etmiss\right)$\ between this 
projected energy fluctuation and the missing transverse momentum
provides a measure of the contribution of the jet to the missing transverse 
energy. The scaled missing transverse energy defined as
\begin{equation}
\etmisspar{Sc} =
\frac{\etmiss}{\sqrt{\sum_{\rm jets}\left(\Delta E^{\rm
        jet}\cdot\sin\theta^{\rm jet}\cdot\cos\Delta\phi\left({\rm
          jet},\etmiss\right)\right)^{2}}}
\end{equation}
is required to be greater than 15 (Cut 3).

The charged lepton system and the neutrinos are emitted
mostly back--to--back, so the invariant mass for the leptons from the Higgs 
decay is restricted to \mh/2.
Thus, the invariant mass \mll\ is required to be $\mll < \mh/2$\ (Cut 4). In 
the $ee$ channel the cut is altered to $\mee < min(80\,{\rm GeV},~\mh/2)$.
In the $\mu\mu$\ channel a lower cut boundary with \mmm\,$>$\,20\,\GeV\ is 
required to remove events from $J/\psi$, $\Upsilon$\ and $Z/\gamma^*$\ 
production.
The sum of the \pt\ of the leptons and \etmiss\ is required to be in the range 
$\mh/2+20\,{\rm GeV}<\ptelf+\ptels+\etmiss<\mh$\ for the $ee$ and $e\mu$\ 
channel and $\mh/2+10\,{\rm GeV}<\ptelf+\ptels+\etmiss<\mh$\ for the $\mu\mu$\ 
channel (Cut 5).
The transverse mass, defined as 
$m_{T}^{\ell\ell'}=\sqrt{2 p_{T}^{\ell\ell'}\etmiss(1-\cos{\Delta\phi(p_{T}^{\ell\ell'},\etmiss)})}$,
with the di-lepton transverse momentum $p_{T}^{\ell\ell'}$, should be in the 
range $\mh/2< m_{T}^{\ell\ell'} < \mh-10\,{\rm GeV}$\ (Cut 6).
The latter two cuts reject events from $W$+jet/$\gamma$ and $WW$ production
and further reduce backgrounds from $Z/\gamma^*$ production.
Finally, to suppress 
the background from \ttbar\ production, the scalar sum of the transverse 
energies of all jets with \Etj\ $>$ 20~\GeV\ and $|\eta| < 2.5$, $H_T$,  is 
required to be less than 100~\GeV\ (Cut 7). 
Remaining $Z$ boson and multi-jet events can be rejected with a cut on the 
opening angle, $\Delta \phi_{\ell\ell'} < 2.0$ (Cut 8),
since most of the backgrounds exhibit a back--to--back topology. This is not 
the case for Higgs boson decays because of the spin correlations in the decay. 
Figure~\ref{fig:dphi_plots} shows the  distributions of the
azimuthal opening angle $\Delta \phi_{\ell\ell'}$ between the two leptons for 
the $ee$\ (b), the $\mu\mu$\ (d) and the $e\mu$ channel (f) 
before applying the final cut on $\Delta \phi_{\ell\ell'}$.

To maximize the sensitivity, the selection in the $\mu\mu$ channel 
is slightly changed for Higgs boson masses \mh = 140 and 160 GeV.
For a better \zg\ background suppression cuts 4, 5 and 6 are replaced by the 
following cuts:
the invariant mass $m_{\mu\mu}$ should be in the range 
$20\,\GeV < m_{\mu\mu} < 80\,\GeV$ (Cut 4). 
Since the momentum resolution is degraded for high \pt\ tracks, 
an additional constrained fit is performed to reject events  
compatible with $Z$ boson production (Cut 5). The sum of the muon transverse 
momenta and the missing transverse energy should be 
$p_{\rm T}^{\mu1}+p_{\rm T}^{\mu2}+\etmiss>90\,{\rm GeV}$\ (Cut 6).

%
The efficiency for \hwwll\ signal events to pass the acceptance and selection
criteria is determined using the {\sc pythia} 6.2 \cite{pythia} event 
generator followed by a detailed {\sc geant}-based \cite{geant}
simulation of the D\O\ detector. All trigger, reconstruction and identification
efficiencies are derived from the data. The kinematic acceptance 
efficiency is derived from MC. The overall detection
efficiencies range from ($0.44\pm0.03$)\% to ($3.9\pm0.2$)\% depending
on the decay channel and \mh. Table~\ref{tab:eff}
summarizes these efficiencies.

\begin{table}[t]
\caption[Efficiency]{\label{tab:eff}Overall detection efficiencies (in \%) 
  for \hwwll\ decays for the three  channels after 
  all cuts. Quoted are the overall uncertainties, combining statistical and 
  systematic components in quadrature. }

\begin{ruledtabular}
\begin{tabular}{lccc} 
\mh\ (\GeV)   & $ee$ & $e\mu$ & $\mu\mu$ \\
\hline
 100  & $0.56\pm0.05$ & $1.02\pm0.06$ & $0.44\pm0.03$ \\
 120  & $1.18\pm0.09$ & $2.0\pm0.1$  & $1.02\pm0.06$ \\
 140  & $1.55\pm0.08$ & $2.9\pm0.2$  & $1.34\pm0.08$ \\
 160  & $2.1\pm0.1$  &  $3.9\pm0.2$ &  $2.0\pm0.1$ \\
 180  & $2.1\pm0.1$  &  $3.9\pm0.2$  & $1.68\pm0.09$ \\
 200  & $1.57\pm0.09$ & $3.2\pm0.1$  & $1.53\pm0.07$ \\
\end{tabular}
\end{ruledtabular}
\end{table}
%
Using the NLO cross sections calculated 
with {\sc higlu}\,\cite{bib:higlu} and 
{\sc hdecay}\,\cite{bib:hdecay}
and the branching ratio $BR$ of $0.1068\pm 0.0012$ for
$W\to \ell\nu$ \cite{pdg}, the expected
number of events for \hww\ decays from all three channels is
$0.68\pm \pm 0.03\,(\rm syst) \pm 0.04\,(\rm lum) $ events for 
a Higgs boson mass \mh $=$ 160~GeV.
The signal expectation for different Higgs masses \mh\ are
given in  Table~\ref{tab:events}.

Background contributions from \zg, $W$+jet/$\gamma$, \ttbar, $WW$, $WZ$ and 
$ZZ$ events are estimated using the {\sc pythia} event generator normalized
to their NLO cross sections\,\cite{wwxsec}. In addition, 
$W$+jet/$\gamma$ contributions are verified using {\sc alpgen} \cite{alpgen}.
All events are processed through the full detector simulation.
The background due to multi-jet production, when a jet is misidentified as an
electron, is determined from the data using a sample of like-sign
di-lepton events with inverted lepton quality cuts (called ``QCD''
in Fig.~\ref{fig:dphi_plots}).
A summary of the background contributions together with signal
expectations and events observed in the data after the final
selection is shown in Table~\ref{tab:events}.
There is good agreement between the number of events
observed in the data and the various backgrounds in all three channels.
The largest difference between the data and the background expectation, 
at \mh = 120 GeV, corresponds to a background probability of 6\%.
The $e\mu$ channel has both
the highest signal efficiency and best signal-to-background ratio.

\begin{table*}[t]
\caption{\label{tab:events} Number of signal and background events expected
  and number of events observed after all
  selections are applied. Only statistical uncertainties are given.}
\begin{ruledtabular}
\begin{tabular}{lcccccc}
 \mh (\GeV)      &   100 & 120 & 140 & 160 & 180 & 200 \\
\hline
\hww             & $0.007\pm0.001$ & $0.125\pm0.002$ & $0.398\pm0.008$  & $0.68\pm0.01$  & $0.463\pm0.009$  & $0.210\pm0.004$ \\       
\hline
\zg              & $7.9\pm 1.1$   & $7.5\pm 1.0$   & $3.8\pm 0.6$   & $4.0\pm 0.7$   & $6.6\pm0.9$    & $9.9\pm1.1$    \\
Diboson          & $4.4\pm 0.2$   & $8.1\pm 0.2$   & $11.7\pm 0.3$  & $12.3\pm 0.3$  & $11.6\pm 0.3$  & $9.6\pm 0.3$   \\
\ttbar           & $0.03\pm 0.01$ & $0.11\pm 0.02$ & $0.29\pm 0.02$ & $0.47\pm 0.03$ & $0.66\pm 0.05$ & $0.72\pm 0.05$ \\
$W$+jet/$\gamma$ & $16.9\pm 2.2$  & $14.2\pm 2.1$  & $5.8\pm 1.2$   & $2.8\pm 0.9$   & $0.7\pm 0.5$   & $0.7\pm 0.5$    \\
Multi-jet         & $0.6\pm0.3$    & $0.3\pm 0.1$   & $0.2\pm 0.1$   & $0.2\pm0.1$    & $0.3\pm 0.1$   & $0.3\pm0.1$    \\             
\hline
Background sum   & $29.9\pm 2.5$  & $30.1\pm 2.3$  & $21.8\pm 1.4$  & $19.7\pm 1.2$  & $19.8\pm 1.1$  & $21.2\pm 1.2$ \\ 
\hline
Data             & 27             & 21             &  20            & 19             & 19             & 14 \\
\end{tabular}
\end{ruledtabular}
\end{table*}
%
\begin{figure}[t]
\includegraphics[width=0.98\columnwidth]{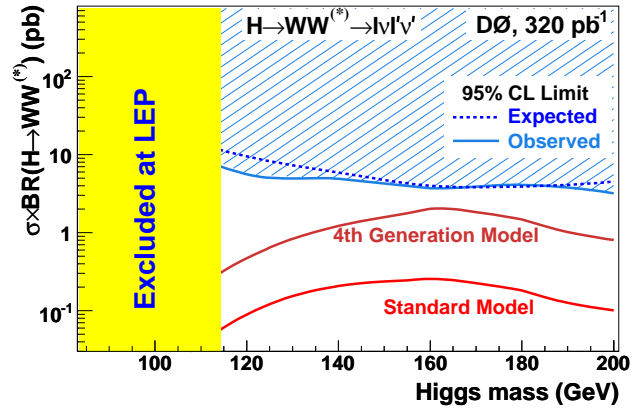}\hfill
 \caption{\label{fig:alllimit} Expected and observed upper limits on the 
   cross section times branching ratio 
   $\sigma\times BR$(\hww) at the 95\% CL together with
   expectations from standard model Higgs boson production and an alternative
   model.
   The {\sc lep} limit on the standard model Higgs boson production is taken 
   from\,\cite{bib:lephiggs} and the 4th generation model prediction is
   described in\,\cite{Arik:2001iw}.
  }
\end{figure}
%
Various sources of systematic uncertainties that affect the background 
estimation and signal efficiencies have been studied.
In these calculations, parameters are varied within $\pm 1\sigma$ of their 
nominal values, where $\sigma$ is determined by the corresponding 
uncertainties.
The trigger efficiency, electron and muon identification efficiencies, 
jet energy scale,  electron and muon momentum resolution, parton distribution 
function uncertainty and cross section calculation of \zg, $WW$ and 
\ttbar\ events contribute to the systematic uncertainties. 
The total systematic uncertainties for the background 
estimate and signal efficiencies for the six Higgs boson masses are given in 
Table~\ref{tab:syst}. The largest contribution to the systematic uncertainty 
on the background for small Higgs boson masses comes from the jet energy scale 
due to the large $W$+jet/$\gamma$ background, whereas for high Higgs boson 
masses the $W$ boson pair production cross section gives the largest systematic
uncertainty. The uncertainty of the parton distribution function is the
largest uncertainty of the signal efficiency.
The uncertainty of the luminosity measurement is 6.5\%.

\begin{table}[t]
\caption{\label{tab:syst} Systematic uncertainties (in \%) of the
  \hww\ signal efficiencies and of the number of background events (BG), for
  the $ee$, $e\mu$ and $\mu\mu$ channels. Uncertainties exclude the 
  uncertainty of the luminosity measurement.}
\begin{ruledtabular}
  \begin{tabular}{lcccccc} 
    & \multicolumn{2}{c}{$ee$} & \multicolumn{2}{c}{$e\mu$} & \multicolumn{2}{c}{$\mu\mu$} \\
    \mh (\GeV)  &  Signal & BG  & Signal & BG  & Signal  & BG    \\
    
    \hline
    $100$ & $8.3$ & $9.5$ & $6.4$ & $11.4$  & $7.8$ & $7.2$  \\
    $120$ & $8.3$ & $8.6$ & $6.7$ & $13.6$  & $7.3$ & $7.5$  \\
    $140$ & $6.4$ & $6.7$ & $6.9$ & $13.6$  & $7.2$ & $8.3$  \\
    $160$ & $6.6$ & $7.3$ & $6.7$ & $12.0$  & $7.1$ & $8.3$  \\
    $180$ & $6.9$ & $10.3$ & $6.6$ & $13.0$  & $7.3$ & $14.6$ \\
    $200$ & $6.8$ & $10.6$ & $6.1$ & $12.3$  & $6.9$ & $18.1$ \\
  \end{tabular}
\end{ruledtabular}
\end{table}
%
Since the remaining candidate events after the selection are consistent 
with the background expectation, limits on the production cross section times 
branching ratio $\sigma \times BR(\hww)$ are derived using a modified
frequentist method described in Ref.\,\cite{bib:junk}. It provides the 
confidence level for the background to
represent the data, $CL_B$, and the confidence level for the sum of signal and
background hypothesis $CL_{S+B}$. The 95\% CL limit is obtained by requiring
$CL_{S+B}/CL_B=0.05$. The uncertainty on the background and the expected
signal events were determined from the statistical and systematic
uncertainties and luminosity uncertainty.
Table~\ref{tab:alllimit} shows the individual expected and observed upper 
limits on the cross section times branching ratio $\sigma \times {BR(\hww)}$
for the combination of the three different decay channels for six different 
Higgs boson masses. The different values of the upper limits are 
due to different background expectations and signal efficiencies. 
The best limits are achieved for 
large Higgs masses since background expectations decrease while signal
efficiencies increase. 

\begin{table}[t]
\caption{\label{tab:alllimit} Expected and observed upper limits at the
  95\% CL for the Higgs boson production cross section times branching
  ratio $\sigma \times {BR(\hww)}$ 
  for six values of \mh. }
\begin{ruledtabular}
\begin{tabular}{lcccccc} 
\mh\ (\GeV) & $100$ & $120$ & $140$ & $160$ & $180$ & $200$ \\  
\hline  
Expected limits (pb) & $20.3$ &  $9.5$ &  $5.9$ &  $4.0$ &  $3.9$ &  $4.5$ \\
Observed limits (pb) & $18.5$ &  $5.6$ &  $4.9$ &  $3.7$ &  $4.1$ &  $3.2$ \\
\end{tabular}
\end{ruledtabular}
\end{table}
%
Figure~\ref{fig:alllimit} shows the expected and observed 
cross section limits for $\sigma\times BR$(\hww) for the different Higgs boson 
masses compared with predictions from the SM and from an extension 
including a fourth fermion family~\cite{Arik:2001iw}.  
With the current dataset, no region of the SM prediction can be excluded. 

To conclude, we have searched for the Higgs boson in \hwwll\ 
($\ell,\ell'=e,\mu,\tau$) decays  with \ee, \emu\ or \mumu final states
in \ppbar\ collisions at $\sqrt{s} = 1.96$ TeV.
The data is consistent with the expectation from backgrounds.
Since no excess has been observed, limits on the production cross section times
branching ratio $\sigma \times BR(\hww)$ have been derived. 

\input acknowledgement_paragraph_r2.tex   


\end{document}

%% file: globals_d0.tex
\newcommand{\mh}{\mbox{${M}_{H}$}}

\newcommand{\mee}{\mbox{${m}_{ee}$}}
\newcommand{\mll}{\mbox{${m}_{\ell\ell}$}}

\newcommand{\mmm}{\mbox{${m}_{\mu\mu}$}}

\newcommand{\pt}{\mbox{$p_{T}$}}

\newcommand{\Etj}{\mbox{$E_{T}^{\text{jet}}$}}

\newcommand{\ptelf}{\mbox{$p_{T}^{\ell_{1}}$}}
\newcommand{\ptels}{\mbox{$p_{T}^{\ell_{2}}$}}

\newcommand{\ipb}{\mbox{$\text{pb}^{-1}$}}

\newcommand{\Eslash}{\mbox{$E \kern-0.6em\slash$}}
\newcommand{\etmiss}{\mbox{$\Eslash_T$}}

\newcommand{\pslash}{\mbox{$p \kern-0.5em\slash$}}

\newcommand{\etmisspar}[1]{\mbox{$\Eslash_T^{\text{#1}}$}}

\newcommand{\hww}{\mbox{$H\rightarrow WW^{(\ast)}$}}

\newcommand{\hwwll}{\mbox{$H\rightarrow WW^{(\ast)}\rightarrow\ell\nu\,{\ell'}\nu'$}}
\newcommand{\zg}{\mbox{$Z/\gamma^*$}}

\newcommand{\ttbar}{\mbox{$t\bar{t}$}}
\newcommand{\ppbar}{\mbox{$p\bar{p}$}}

\newcommand{\ee}{\mbox{$e^{+}e^{-}$}}
\newcommand{\emu}{\mbox{$e^{\pm}\mu^{\mp}$}}
\newcommand{\mumu}{\mbox{$\mu^{+}\mu^{-}$}}

\newcommand{\GeV}{\text{GeV}}


%% file: list_of_authors_r2.tex
%
\author{                                                                      
V.M.~Abazov,$^{35}$                                                           
B.~Abbott,$^{72}$                                                             
M.~Abolins,$^{63}$                                                            
B.S.~Acharya,$^{29}$                                                          
M.~Adams,$^{50}$                                                              
T.~Adams,$^{48}$                                                              
M.~Agelou,$^{18}$                                                             
J.-L.~Agram,$^{19}$                                                           
S.H.~Ahn,$^{31}$                                                              
M.~Ahsan,$^{57}$
% ...
\\                                                                            
\vskip 0.30cm                                                                 
\centerline{(D\O\ Collaboration)}                                             
\vskip 0.30cm                                                                 
}                                                                             
\affiliation{                                                                 
\centerline{$^{1}$Universidad de Buenos Aires, Buenos Aires, Argentina}       
\centerline{$^{2}$LAFEX, Centro Brasileiro de Pesquisas F{\'\i}sicas,         
                  Rio de Janeiro, Brazil}                                     
\centerline{$^{3}$Universidade do Estado do Rio de Janeiro,                   
                  Rio de Janeiro, Brazil}                                     
\centerline{$^{4}$Instituto de F\'{\i}sica Te\'orica, Universidade            
                  Estadual Paulista, S\~ao Paulo, Brazil}                     
\centerline{$^{5}$University of Alberta, Edmonton, Alberta, Canada,           
               Simon Fraser University, Burnaby, British Columbia, Canada,}   
\centerline{York University, Toronto, Ontario, Canada, and                    
         McGill University, Montreal, Quebec, Canada}                         
\centerline{$^{6}$Institute of High Energy Physics, Beijing,                  
                  People's Republic of China}                                 
\centerline{$^{7}$University of Science and Technology of China, Hefei,       
                  People's Republic of China}                                 
\centerline{$^{8}$Universidad de los Andes, Bogot\'{a}, Colombia}             
\centerline{$^{9}$Center for Particle Physics, Charles University,            
                  Prague, Czech Republic}                                     
\centerline{$^{10}$Czech Technical University, Prague, Czech Republic}        
\centerline{$^{11}$Center for Particle Physics, Institute of Physics,         
                   Academy of Sciences of the Czech Republic,                 
                   Prague, Czech Republic}                                    
\centerline{$^{12}$Universidad San Francisco de Quito, Quito, Ecuador}        
\centerline{$^{13}$Laboratoire de Physique Corpusculaire, IN2P3-CNRS,         
                  Universit\'e Blaise Pascal, Clermont-Ferrand, France}       
\centerline{$^{14}$Laboratoire de Physique Subatomique et de Cosmologie,      
                  IN2P3-CNRS, Universite de Grenoble 1, Grenoble, France}     
\centerline{$^{15}$CPPM, IN2P3-CNRS, Universit\'e de la M\'editerran\'ee,     
                  Marseille, France}                                          
\centerline{$^{16}$IN2P3-CNRS, Laboratoire de l'Acc\'el\'erateur              
                  Lin\'eaire, Orsay, France}                                  
\centerline{$^{17}$LPNHE, IN2P3-CNRS, Universit\'es Paris VI and VII,         
                  Paris, France}                                              
\centerline{$^{18}$DAPNIA/Service de Physique des Particules, CEA, Saclay,    
                  France}                                                     
\centerline{$^{19}$IReS, IN2P3-CNRS, Universit\'e Louis Pasteur, Strasbourg,  
                France, and Universit\'e de Haute Alsace, Mulhouse, France}   
\centerline{$^{20}$Institut de Physique Nucl\'eaire de Lyon, IN2P3-CNRS,      
                   Universit\'e Claude Bernard, Villeurbanne, France}         
\centerline{$^{21}$III. Physikalisches Institut A, RWTH Aachen,               
                   Aachen, Germany}                                           
\centerline{$^{22}$Physikalisches Institut, Universit{\"a}t Bonn,             
                  Bonn, Germany}                                              
\centerline{$^{23}$Physikalisches Institut, Universit{\"a}t Freiburg,         
                  Freiburg, Germany}                                          
\centerline{$^{24}$Institut f{\"u}r Physik, Universit{\"a}t Mainz,            
                  Mainz, Germany}                                             
\centerline{$^{25}$Ludwig-Maximilians-Universit{\"a}t M{\"u}nchen,            
                   M{\"u}nchen, Germany}                                      
\centerline{$^{26}$Fachbereich Physik, University of Wuppertal,               
                   Wuppertal, Germany}                                        
\centerline{$^{27}$Panjab University, Chandigarh, India}                      
\centerline{$^{28}$Delhi University, Delhi, India}                            
\centerline{$^{29}$Tata Institute of Fundamental Research, Mumbai, India}     
\centerline{$^{30}$University College Dublin, Dublin, Ireland}                
\centerline{$^{31}$Korea Detector Laboratory, Korea University,               
                   Seoul, Korea}                                              
\centerline{$^{32}$CINVESTAV, Mexico City, Mexico}                            
\centerline{$^{33}$FOM-Institute NIKHEF and University of                     
                  Amsterdam/NIKHEF, Amsterdam, The Netherlands}               
\centerline{$^{34}$Radboud University Nijmegen/NIKHEF, Nijmegen, The          
                  Netherlands}                                                
\centerline{$^{35}$Joint Institute for Nuclear Research, Dubna, Russia}       
\centerline{$^{36}$Institute for Theoretical and Experimental Physics,        
                  Moscow, Russia}                                             
\centerline{$^{37}$Moscow State University, Moscow, Russia}                   
\centerline{$^{38}$Institute for High Energy Physics, Protvino, Russia}       
\centerline{$^{39}$Petersburg Nuclear Physics Institute,                      
                   St. Petersburg, Russia}                                    
\centerline{$^{40}$Lund University, Lund, Sweden, Royal Institute of          
                   Technology and Stockholm University, Stockholm,            
                   Sweden, and}                                               
\centerline{Uppsala University, Uppsala, Sweden}                              
\centerline{$^{41}$Lancaster University, Lancaster, United Kingdom}           
\centerline{$^{42}$Imperial College, London, United Kingdom}                  
\centerline{$^{43}$University of Manchester, Manchester, United Kingdom}      
\centerline{$^{44}$University of Arizona, Tucson, Arizona 85721, USA}         
\centerline{$^{45}$Lawrence Berkeley National Laboratory and University of    
                  California, Berkeley, California 94720, USA}                
\centerline{$^{46}$California State University, Fresno, California 93740, USA}
\centerline{$^{47}$University of California, Riverside, California 92521, USA}
\centerline{$^{48}$Florida State University, Tallahassee, Florida 32306, USA} 
\centerline{$^{49}$Fermi National Accelerator Laboratory, Batavia,            
                   Illinois 60510, USA}                                       
\centerline{$^{50}$University of Illinois at Chicago, Chicago,                
                   Illinois 60607, USA}                                       
\centerline{$^{51}$Northern Illinois University, DeKalb, Illinois 60115, USA} 
\centerline{$^{52}$Northwestern University, Evanston, Illinois 60208, USA}    
\centerline{$^{53}$Indiana University, Bloomington, Indiana 47405, USA}       
\centerline{$^{54}$University of Notre Dame, Notre Dame, Indiana 46556, USA}  
\centerline{$^{55}$Iowa State University, Ames, Iowa 50011, USA}              
\centerline{$^{56}$University of Kansas, Lawrence, Kansas 66045, USA}         
\centerline{$^{57}$Kansas State University, Manhattan, Kansas 66506, USA}     
\centerline{$^{58}$Louisiana Tech University, Ruston, Louisiana 71272, USA}   
\centerline{$^{59}$University of Maryland, College Park, Maryland 20742, USA} 
\centerline{$^{60}$Boston University, Boston, Massachusetts 02215, USA}       
\centerline{$^{61}$Northeastern University, Boston, Massachusetts 02115, USA} 
\centerline{$^{62}$University of Michigan, Ann Arbor, Michigan 48109, USA}    
\centerline{$^{63}$Michigan State University, East Lansing, Michigan 48824,   
                   USA}                                                       
\centerline{$^{64}$University of Mississippi, University, Mississippi 38677,  
                   USA}                                                       
\centerline{$^{65}$University of Nebraska, Lincoln, Nebraska 68588, USA}      
\centerline{$^{66}$Princeton University, Princeton, New Jersey 08544, USA}    
\centerline{$^{67}$Columbia University, New York, New York 10027, USA}        
\centerline{$^{68}$University of Rochester, Rochester, New York 14627, USA}   
\centerline{$^{69}$State University of New York, Stony Brook,                 
                   New York 11794, USA}                                       
\centerline{$^{70}$Brookhaven National Laboratory, Upton, New York 11973, USA}
\centerline{$^{71}$Langston University, Langston, Oklahoma 73050, USA}        
\centerline{$^{72}$University of Oklahoma, Norman, Oklahoma 73019, USA}       
\centerline{$^{73}$Oklahoma State University, Stillwater, Oklahoma 74078, USA}
\centerline{$^{74}$Brown University, Providence, Rhode Island 02912, USA}     
\centerline{$^{75}$University of Texas, Arlington, Texas 76019, USA}          
\centerline{$^{76}$Southern Methodist University, Dallas, Texas 75275, USA}   
\centerline{$^{77}$Rice University, Houston, Texas 77005, USA}                
\centerline{$^{78}$University of Virginia, Charlottesville, Virginia 22901,   
                   USA}                                                       
\centerline{$^{79}$University of Washington, Seattle, Washington 98195, USA}  
}                                                                             

%% file: acknowledgement_paragraph_r2.tex
%
We thank the staffs at Fermilab and collaborating institutions, 
and acknowledge support from the 
DOE and NSF (USA);
CEA and CNRS/IN2P3 (France);
FASI, Rosatom and RFBR (Russia);
CAPES, CNPq, FAPERJ, FAPESP and FUNDUNESP (Brazil);
DAE and DST (India);
Colciencias (Colombia);
CONACyT (Mexico);
KRF (Korea);
CONICET and UBACyT (Argentina);
FOM (The Netherlands);
PPARC (United Kingdom);
MSMT (Czech Republic);
CRC Program, CFI, NSERC and WestGrid Project (Canada);
BMBF and DFG (Germany);
SFI (Ireland);
Research Corporation,
Alexander von Humboldt Foundation,
and the Marie Curie Program.